\pdfoutput=1

\documentclass[11pt,twoside,a4paper,cmspaper,final,collab]{cms-tdr}
\def\svnVersion{360578}\def\cmsCernNoTag{CERN-EP/2016-012}\def\cmsCernDate{\today}\def\cmsMessage{Published in Physics Letters B as \href{http://dx.doi.org/10.1016/j.physletb.2016.05.088}{\doi{10.1016/j.physletb.2016.05.088.}}}

\begin{document}\cmsNoteHeader{SUS-14-016}

\hyphenation{had-ron-i-za-tion}
\hyphenation{cal-or-i-me-ter}
\hyphenation{de-vices}
\RCS$Revision: 341520 $
\RCS$HeadURL: svn+ssh://svn.cern.ch/reps/tdr2/papers/SUS-14-016/trunk/SUS-14-016.tex $
\RCS$Id: SUS-14-016.tex 341520 2016-05-04 19:04:38Z alverson $
\newlength\cmsFigWidth
\ifthenelse{\boolean{cms@external}}{\setlength\cmsFigWidth{0.85\columnwidth}}{\setlength\cmsFigWidth{0.4\textwidth}}
\ifthenelse{\boolean{cms@external}}{\providecommand{\cmsLeft}{top\xspace}}{\providecommand{\cmsLeft}{left\xspace}}
\ifthenelse{\boolean{cms@external}}{\providecommand{\cmsRight}{bottom\xspace}}{\providecommand{\cmsRight}{right\xspace}}
\cmsNoteHeader{SUS-14-016}
\title{Search for supersymmetry in electroweak production with photons and large missing transverse energy in pp collisions at $\sqrt{s}=8$\TeV}

\date{\today}

\abstract{
Results are reported from a search for supersymmetry with gauge-mediated supersymmetry breaking in electroweak production. Final states with photons and large missing transverse energy (\MET) were examined. The data sample was collected in pp collisions at $\sqrt{s}=8\TeV$ with the CMS detector at the LHC and corresponds to 7.4\fbinv. The analysis focuses on scenarios in which the lightest neutralino has bino- or wino-like components, resulting in decays to photons and gravitinos, where the gravitinos escape undetected. The data were obtained using a specially designed trigger with dedicated low thresholds, providing good sensitivity to signatures with photons, \MET, and low hadronic energy. No excess of events over the standard model expectation is observed. The results are interpreted using the model of general gauge mediation. With the wino mass fixed at 10\GeV above that of the bino, wino masses below 710\GeV are excluded at 95\% confidence level. Constraints are also set in the context of two simplified models, for which the analysis sets the lowest cross section limits on the electroweak production of supersymmetric particles.
}

\newcommand{\lumiPhoton}{\ensuremath{7.4\fbinv}\xspace}
\newcommand{\metsig}{\ensuremath{E_{\mathrm{T}}^{\text{miss,signif}} }\xspace}
\newcommand{\JSigma}{\ensuremath{S_{\mathrm{T}}^{\gamma} }\xspace}
\newcommand{\ptg}{\ensuremath{p_{\mathrm{T}}^{\gamma_{1}} }\xspace}
\newcommand{\ptgi}{\ensuremath{p_{\mathrm{T}}^{\gamma_{i}} }\xspace}
\providecommand{\PV}{\ensuremath{\mathrm{V}}\xspace}
\newcommand{\Vgamma}{\ensuremath{\PV\gamma}\xspace}
\newcommand{\gammaJets}{\ensuremath{\gamma\text{+jets} }\xspace}
\newcommand{\TchiNglimit}{\ensuremath{570\GeV}\xspace}
\newcommand{\TchiWglimit}{\ensuremath{680\GeV}\xspace}
\newcommand{\WinoBinoLimit}{\ensuremath{710\GeV}\xspace}

\hypersetup{%
pdfauthor={CMS Collaboration},%
pdftitle={Search for supersymmetry in electroweak production with photons and large missing transverse energy in pp collisions at sqrt(s)=8 TeV},%
pdfsubject={CMS},%
pdfkeywords={CMS, physics, supersymmetry}}

\maketitle

\section{Introduction}
\label{sec:introduction}

Supersymmetry~\cite{Ramond,Ramond:1971kx,Golfand,Volkov,Wess:1974tw,Freedman:1976xh,Deser:1976eh,Freedman:1976py,Ferrara:1976fu,Fayet,Chamseddine,Barbieri,Hall,Kane} (SUSY) can stabilize the mass of the Higgs boson, recently measured to be around 125~GeV~\cite{Chatrchyan:2012xdj,Aad:2012tfa}, and hence the electroweak scale against large quantum corrections, thus providing a solution to the gauge hierarchy problem~\cite{Barbieri198863}. Searches for supersymmetric partners of standard model (SM) particles with photons in the final state are already probing SUSY parameter space at the \TeV scale~\cite{Chatrchyan:2011wc,Chatrchyan:2012bba,Khachatryan:2015exa,Aad:2015hea}. There is a strong interest in probing so-called natural SUSY scenarios, where a subset of SUSY partners can remain light, while many other SUSY partners can have large masses that are inaccessible to present searches. These regions of SUSY parameter space are still largely unexplored.

In this analysis, $R$-parity~\cite{Barbier:2004ez,rparity} is assumed to be conserved, so that SUSY particles are always produced in pairs. In SUSY models of gauge-mediated SUSY breaking (GMSB)~\cite{GGMa,GGMd2,GGMd3,GGMd4,GGMd5,GGMd1,GGMd} the gravitino (\sGra) is the lightest SUSY particle (LSP) and escapes undetected, leading to missing transverse energy (\MET) in the detector. In the studied cases, the next-to-lightest SUSY particle (NLSP) is the lightest neutralino ($\chiz_1$). Depending on its composition, the $\chiz_1$ can decay according to $\chiz_1\to N\sGra$, where $N$ is either a photon $\gamma$, a SM Higgs boson H, or a \Z boson. If the gauginos are nearly mass-degenerate, chargino ($\chipm_1$) decays according to $\chipm_1\to \Wpm\sGra$ are also possible.

The ATLAS and CMS collaborations have searched for direct electroweak production of gauginos. Final states with at least one photon and one electron or muon have been examined~\cite{CMS:2015loa,Aad:2015hea}, requiring one
gaugino decaying to $\gamma\sGra$ and one to $\Wpm\sGra$. The NLSP masses below 540\GeV in the simplified model spectra TChiWg scenario, introduced below, were excluded at the 95\% confidence level (CL). In decays into any of the heavy standard model
bosons (\PH, \Z, \Wpm), higgsino (chargino) masses up to 380\GeV (210\GeV) have been excluded at the 95\% CL~\cite{Khachatryan:2014mma,Aad:2015jqa,Aad:2014nua}. Other analyses requiring two photons in the final
state~\cite{Chatrchyan:2012bba,Khachatryan:2015exa,Aad:2015hea} probe bino-like neutralinos and within the context of general gauge mediation
(GGM)~\cite{GGMe,GGMf,Ruderman:2011vv,Kats:2011qh,Kats:2012ym,Grajek:2013ola} exclude electroweakly produced winos with masses up to 740\GeV, depending on the bino mass. A previous single-photon
analysis~\cite{Khachatryan:2015exa} has set limits on bino- and wino-like neutralinos for strong production, but the search is insensitive to electroweak production because the chosen trigger requires $\HT>500$\GeV, where \HT is the scalar sum of transverse energy clustered in jets.

To provide sensitivity to GMSB scenarios with low gaugino masses and mass differences, this analysis uses signatures with at least one photon together with large \MET. In signal events, hadronic energy arises only from initial-state radiation or from the decays of a \Wpm or \Z boson, collectively denoted as \PV~bosons. We concentrate on final states with only a moderate amount of \HT due to direct electroweak production of gauginos. The lightest gauginos are assumed to be either bino- or wino-like, leading to final states with \MET and $\gamma\gamma$, $\gamma\PV$, or $\PV\PV$.

\begin{figure*}[tb]
\centering
\includegraphics[width=0.40\textwidth]{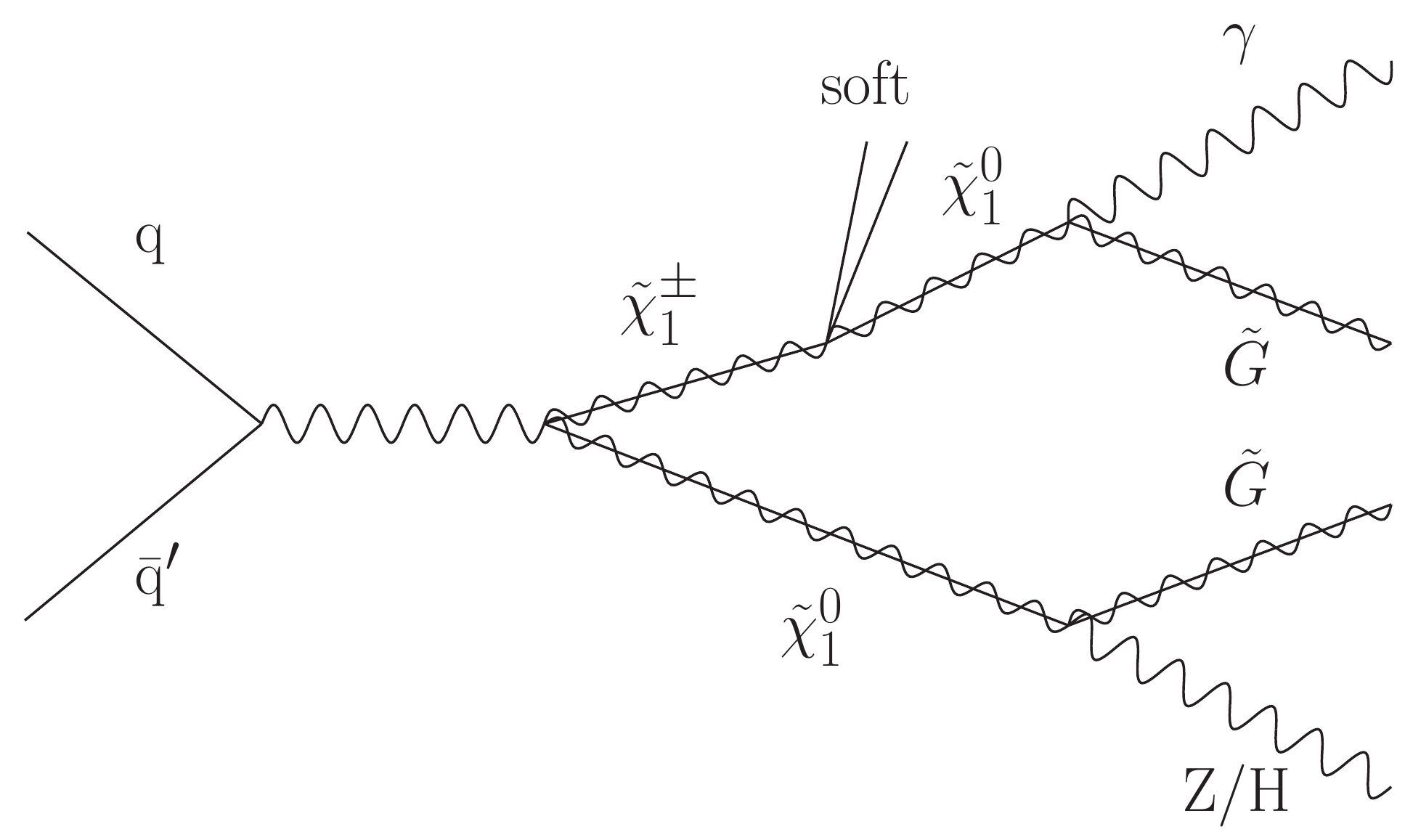}
\includegraphics[width=0.40\textwidth]{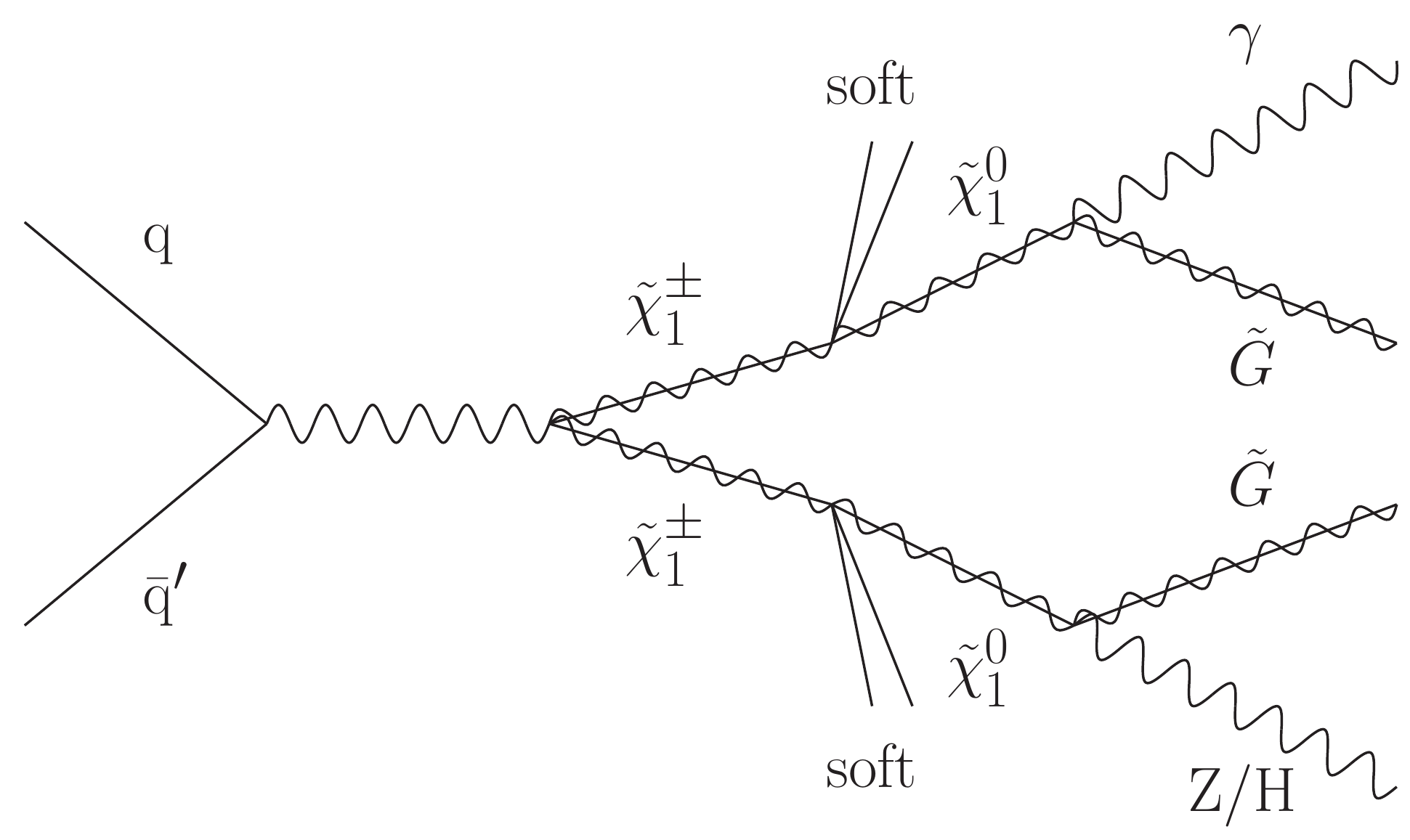}\\
\includegraphics[width=0.40\textwidth]{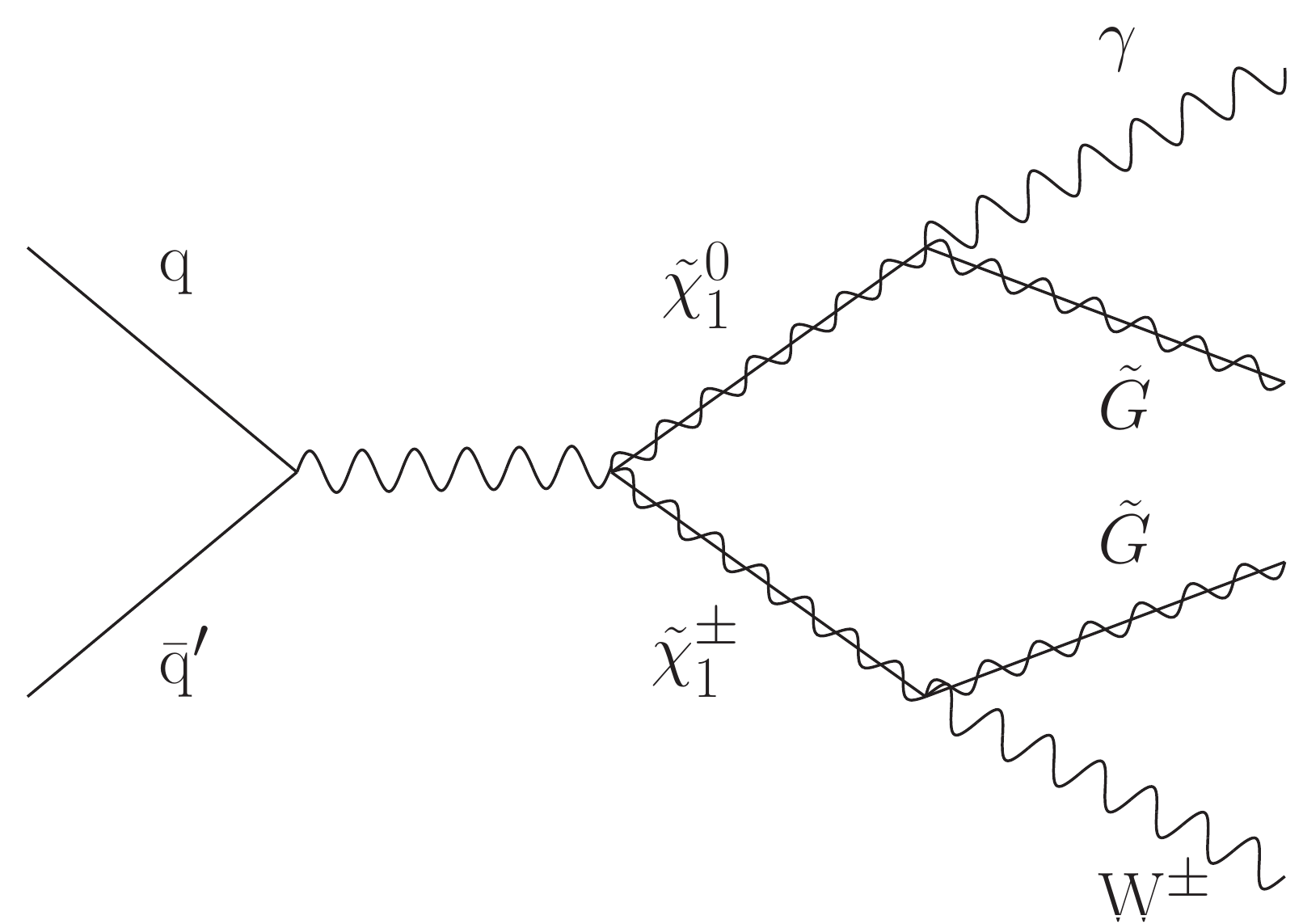}
\includegraphics[width=0.40\textwidth]{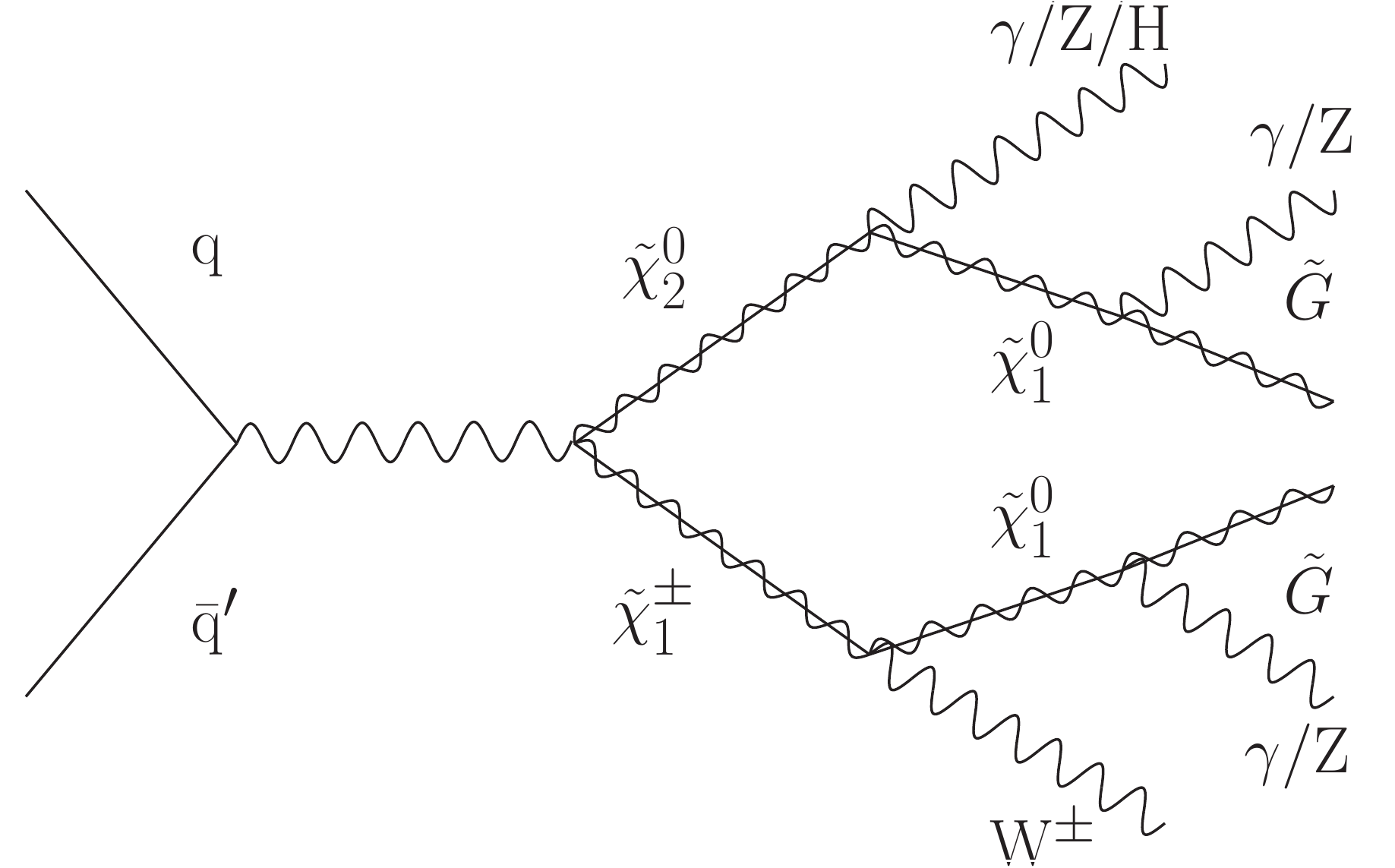}
  \caption{
  Scenarios for the production and decay of charginos and neutralinos considered in this analysis.
  In the TChiNg scenario (top row), the charginos are only slightly heavier than the neutralinos, leading to chargino to neutralino decays accompanied by soft radiation. One neutralino decays to a photon and a gravitino, while the other decays into a \Z or an \PH boson and a gravitino with equal probability. In the TChiWg scenario (bottom left), the gauginos are mass-degenerate and the $\chiz_1$ decays are as shown. Within GGM models, the $\chiz_1\to\gamma\sGra$ to $\chiz_1\to \Z\sGra$ branching fraction depends on the neutralino mass. The dominant process for electroweak GGM production is shown in bottom right. A smaller amount of hadronic energy compared to strong production and at least one photon and \MET are common features of all scenarios.} \label{fig:feynmans}
\end{figure*}

We consider three signal scenarios: The scenario TChiNg models the electroweak pair and associated production of nearly mass-degenerate charginos and neutralinos, which then decay into the NLSP, as shown in Figs.~\ref{fig:feynmans} (top row). The branching fractions of the NLSP decay correspond here to a wino-like $\chiz_1$ of similar mass. The TChiWg scenario models associative production of mass-degenerate charginos and neutralinos, which then decay as shown in Fig.~\ref{fig:feynmans} (bottom left). The third scenario is electroweak production within the GGM context; the dominant production channel is shown in
Fig.~\ref{fig:feynmans} (bottom  right). Masses of the bino- and wino-like neutralinos involved in this scenario are scanned, while the squark and gluino masses are decoupled. The amount of \MET and the photon transverse momentum (\PT) is
determined by the mass of the $\chiz_1$, while the mass of the $\chipm$ determines the production cross section. In the GGM framework, where the gauginos are not mass-degenerate by construction, a larger $\chipm$-$\chiz_1$ mass difference increases the hadronic energy in the final state.

The analysis uses a special data set corresponding to an integrated luminosity of \lumiPhoton recorded with a trigger requiring a photon candidate with $\PT>30\GeV$ measured within $\abs{\eta} < 1.44$ and $\MET>25\GeV$, where the \MET is calculated using calorimeter information without correcting for muons.
Compared to the available triggers in the full 2012 data set, which require a photon candidate with $\PT>135\GeV$ or events with $\MET>120\GeV$, these low trigger thresholds enable a high signal sensitivity to electroweak production and compressed mass spectrum scenarios. The data set was recorded during the second half of the 2012 data-taking period but only reconstructed during the Long Shutdown 1 of the LHC as part of the so-called ``parked-data'' program~\cite{CMS-DP-2012-022}.

\section{The CMS detector}
\label{sec:detector}

The central feature of the CMS apparatus is a superconducting solenoid of 6\unit{m} internal dia\-meter, providing a magnetic field of 3.8\unit{T}. Within the solenoid volume are a silicon pixel and strip tracker, a lead tungstate crystal electromagnetic calorimeter (ECAL), and a brass and scintillator hadron calorimeter (HCAL), each composed of a barrel and two endcap sections. Extensive forward calorimetry complements the coverage provided by the barrel and endcap detectors. Muons are measured in gas-ionization detectors embedded in the steel flux-return yoke outside the solenoid.

In the barrel section of the ECAL, an energy resolution of approximately 1\% is achieved for unconverted or late-converting photons arising from the $\PH\to\Pgg\Pgg$ decay. The remaining barrel photons have an energy resolution of about 1.3\% up to a pseudorapidity of $\abs{\eta}=1$, rising to about 2.5\% at $\abs{\eta}=1.4$~\cite{Khachatryan:2015iwa}.

A more detailed description of the CMS detector, together with a definition of the coordinate system used and the relevant kinematic variables, can be found in Ref.~\cite{Chatrchyan:2008zzk}.

\section{Object reconstruction and simulation}
\label{sec:reconstruction}
Photons are reconstructed~\cite{Khachatryan:2015iwa} from clusters in the ECAL barrel with $\abs{\eta} < 1.44$ and are required to be isolated. The energy deposit in the HCAL tower closest to the seed of the ECAL supercluster assigned to the photon divided by the energy deposit in the ECAL is required to be less than $5\%$. A photon-like shower shape is required. The photon isolation is determined by computing the transverse energy in a cone centered around the photon momentum vector. The cone has an outer radius of 0.3 in $\Delta R = \sqrt{\smash[b]{(\Delta\phi)^2 + (\Delta\eta)^2}}$, where $\phi$ is azimuthal angle in radians, and the contribution of the photon is removed. Corrections for the effects of multiple interactions in the same bunch crossing (pileup) are applied to all isolation energies, depending on the $\eta$ of the photon. Discrimination against electrons is achieved by requiring that photons have no matching pattern of hits in the pixel detector.

All objects used in this search, i.e. photons, electrons, muons, and jets, are reconstructed using the particle-flow (PF) algorithm~\cite{CMS-PAS-PFT-09-001,CMS-PAS-PFT-10-001}.
Jets are reconstructed with the anti-\kt clustering
algorithm~\cite{Cacciari:2008gp} as implemented in the \FASTJET~\cite{fastjet} package, using a distance parameter of 0.5. The jets are required to have transverse momenta above 30\GeV and to satisfy the requirement $\abs{\eta} < 2.4$. Pileup corrections and corrections for the response of the detector are applied to the momenta of the jets~\cite{Cacciari:2007fd,2011JInst...611002C}.

The missing \PT vector is defined as the projection onto the plane perpendicular to the beams of the negative vector sum of the momenta of all reconstructed particles in an event. Its magnitude is referred to as \ETmiss. The \ETmiss is calculated using all particles identified by the PF algorithm. Filters against anomalously high \MET from instrumental effects are applied~\cite{Khachatryan:2014gga}.

The dimensionless variable \MET-significance (\metsig)~\cite{Khachatryan:2014gga} is a measure of the probability that the \MET in a given event arises from genuine
noninteracting stable particles, such as neutrinos or gravitinos, and not as a consequence of the limited energy resolution of objects, such as jets,
photons, or leptons. The \metsig is proportional to the logarithm of the likelihood ratio of both hypotheses. Energy and angular resolutions of the relevant objects, determined by data and simulation
studies, are taken into account. For processes where the reconstructed \MET only arises from the limited energy resolution, the \metsig is an exponentially falling distribution, falling by three orders of magnitude for $0 < \metsig < 25$. SM processes with genuine \MET have approximately the same decrease between 0 and 125. Electroweak SUSY processes intrinsically have high values of genuine \MET along with a significantly lower presence of jets, which otherwise increase the energy uncertainty used in the calculation of \metsig. Therefore, the \metsig distribution of those processes does not decrease monotonically, and a considerable fraction have $\metsig >200$.

The SM $\ttbar\gamma$ and QCD multijet production samples, as well as the TChiNg and TChiWg signal scenarios, are simulated with the \MADGRAPH 5.1.3 generator~\cite{Alwall:2014hca}. The diboson event samples and the electroweak GGM signal scan are generated using \PYTHIA 6.4~\cite{Sjostrand:2006za}. All Monte Carlo (MC) samples incorporate the CTEQ6L1~\cite{Pumplin:2002vw} parton distribution functions (PDF) and use the \PYTHIA program to describe the parton showering and the hadronization. The \GEANTfour~\cite{Agostinelli:2002hh} package is used to model the detector and detector response. The cross sections of the electroweak GGM signal scan are calculated at next-to-leading-order (NLO) accuracy using the \PROSPINO~2.1.~\cite{Beenakker:1996ed} program, the cross sections for the TChiNg and TChiWg signal points are calculated at NLO+NLL (next-to-leading logarithm) accuracy \cite{Kramer:2012bx, Fuks:2012qx, Fuks:2013vua}.

\section{Analysis}
\label{sec:photon}
The data are selected by a trigger with \MET and photon \PT requirements. The events are subsequently required to contain at least one tightly isolated photon measured in the ECAL barrel
with a \PT of at least 40\GeV. The \MET is required to exceed 100\GeV. In addition, \HT is required to exceed 100\GeV, improving the signal-to-background ratio. With this selection the parked data set trigger efficiency is uniform and measured to be $86.5^{+1.0}_{-1.3}\%$.
No selection criteria were applied on the presence or absence of an identified lepton.
All events are required to have $\metsig>80$ and a transverse mass $M_{T}(\MET,\ptg)>300\GeV$, where \ptg is the transverse momentum of the photon with the highest \PT. The cut on \metsig strongly reduces the contribution of backgrounds without genuine \MET, whereas the cut on $M_{T}(\MET,\ptg)$ affects all backgrounds but retains most of the signal events, since only in signal processes the photon and the source for \MET are expected to originate from the same mother particle.
To increase the sensitivity to higher gaugino masses, the signal region is divided into four exclusive bins defined by regions in the variables $\JSigma$ and $\metsig$,
where $\JSigma$ is defined by $\JSigma = \sum_{i} \ptgi + \MET$. The boundaries of the bins are at $\JSigma=600\GeV$ and $\metsig=200$.

The dominant SM backgrounds are vector-boson production with initial- or final-state photon radiation (\Vgamma) and direct photon production (\gammaJets). The normalization of these two backgrounds is determined simultaneously by a $\chi^2$-fit in the control region selection defined by $\metsig>10$ and $M_{T}(\MET,\ptg)>100\GeV$, but excluding the signal region defined above. The distribution of $\MET/\sqrt{\HT}$ is chosen as template variable for the $\chi^2$-fit, which sufficiently separates the shapes of \Vgamma and \gammaJets, so that scaling one background cannot compensate the other. The shape of both backgrounds is simulated with \MADGRAPH 5.1.3. Under the constraint of a fixed total yield, the scale factors for the \Vgamma and \gammaJets simulations are given by the minimum of the $\chi^2$/ndf distribution and found to be $f_{\Vgamma}=0.94\pm0.23$ and $f_{\gammaJets}=2.20\pm0.31$, respectively. Before performing the normalization, the \Vgamma background is scaled to the NLO cross section \cite{Khachatryan:2014rwa}, whereas \gammaJets is used with LO cross section calculated by the event generator. The upper and lower uncertainty is given by the difference of the
best estimate and the scale factor corresponding to the $\chi^2$/ndf values at the minimum of the parabola increased by unity. The measured scale factors and their uncertainties were studied and found stable with respect to systematic variations in the background prediction over
different control regions, template variables, and binnings of the template variables.  The anticorrelation of the \Vgamma and \gammaJets systematic uncertainties due to the fixed total normalization is taken into account in the interpretation. Signal contamination becomes relevant if the gauginos are light because the signal kinematics for light gauginos are similar to that of \Vgamma production. In the examined phase space, signal contamination is negligible.

A subdominant background arises from electrons misidentified as photons ($\Pe\to\gamma$). The misidentification rate $f_{\Pe\to\gamma}=(1.46\pm0.16)\%$ is determined from $\PZ\to\Pep\Pem$ decays in data~\cite{Khachatryan:2015exa}.
The background is estimated from a data control sample with the same event selection, but containing an identified electron instead of a photon. The prediction of electrons misidentified as photons is then obtained by scaling this control sample by $f_{\Pe\to\gamma}/(1-f_{\Pe\to\gamma})$. The uncertainty of this estimation is $11\%$, which is dominated by the misidentification rate uncertainty.
Further minor contributions from $\ttbar\gamma$, diboson, and QCD multijet production are estimated using MC simulations and are corrected for electrons misidentified as photons
at the generator level to avoid overlaps. For the cross sections, 26\%, 50\%, and 100\% systematic uncertainties are assigned to $\ttbar\gamma$, diboson, and QCD multijet backgrounds, respectively. Based on simulation studies, the background from QCD multijet events is found to be negligible.

The systematic uncertainties with respect to the choice of the PDF in the signal acceptance are determined by the difference in acceptance using different sets of PDFs~\cite{Alekhin:2011sk,Botje:2011sn,Ball:2012cx,Martin:2009iq,Lai:2010vv} and vary from less than 1\% to 11\%. Further
systematic uncertainties arise from the jet energy correction (0.1--2.4\% for the signal, 1.3\% for the background estimation) and from the integrated luminosity measurement
(2.6\%)~\cite{CMS-PAS-LUM-13-001}. When evaluating the exclusion contours for SUSY particle masses in specific models, signal cross sections ($\sigma_s$) are conservatively lowered by one standard deviation (4-8\%) corresponding to the combined theoretical uncertainty in $\sigma_s$ due to the choices of the renormalization and factorization scales and the PDFs. All systematic uncertainties are summarized in Table~\ref{tab:syst_uncert}.

\begin{table*}[htb!]
\centering
\topcaption{Summary table of systematic uncertainties relevant for the analysis. Uncertainties due to the luminosity and trigger efficiency measurement apply only to the backgrounds estimated using MC
simulation without data normalization, namely $\ttbar\gamma$, diboson, and multijet, and for the signal. The total uncertainty is dominated by the uncertainty in the $\PV\gamma$ background.}
\label{tab:syst_uncert}
\begin{tabular}{l|l|c|c}
\hline
\multirow{2}{*}{Source} & \multirow{2}{*}{Sample} & \multicolumn{2}{c}{Relative uncert. ( \% )} \\
&  &  in sample & in total bkg \\
\hline
$\PV\gamma$ normalization & $\PV\gamma$ & 24 & 19 \\
$\gamma$+jets normalization & $\gamma$+jets & 14 & 1  \\
$\PZ\to \Pe^+\Pe^-$ fit & $\Pe\to\gamma$ & 11 & 0.3 \\
Cross section measurement & $\ttbar\gamma$ & 26 & 3 \\
Cross section & Diboson & 50& 1 \\
Cross section & Multijet & 100 & 0 \\
Integrated luminosity & Diboson, multijet, and signal & 2.6 & --- \\
Trigger efficiency & Diboson, multijet, and signal & 1.2 & --- \\
Jet energy scale & Diboson, multijet, and signal & $1{-}2$ & --- \\
PDF uncertainty in acceptance & Signal & ${<}1{-}11$ & --- \\
PDF and scale uncertainty & Signal & 4-8 & --- \\
\hline
\end{tabular}
\end{table*}

\section{Results and interpretation}
\label{sec:results}

\begin{figure*}[htb]
\centering
\includegraphics[width=0.49\textwidth]{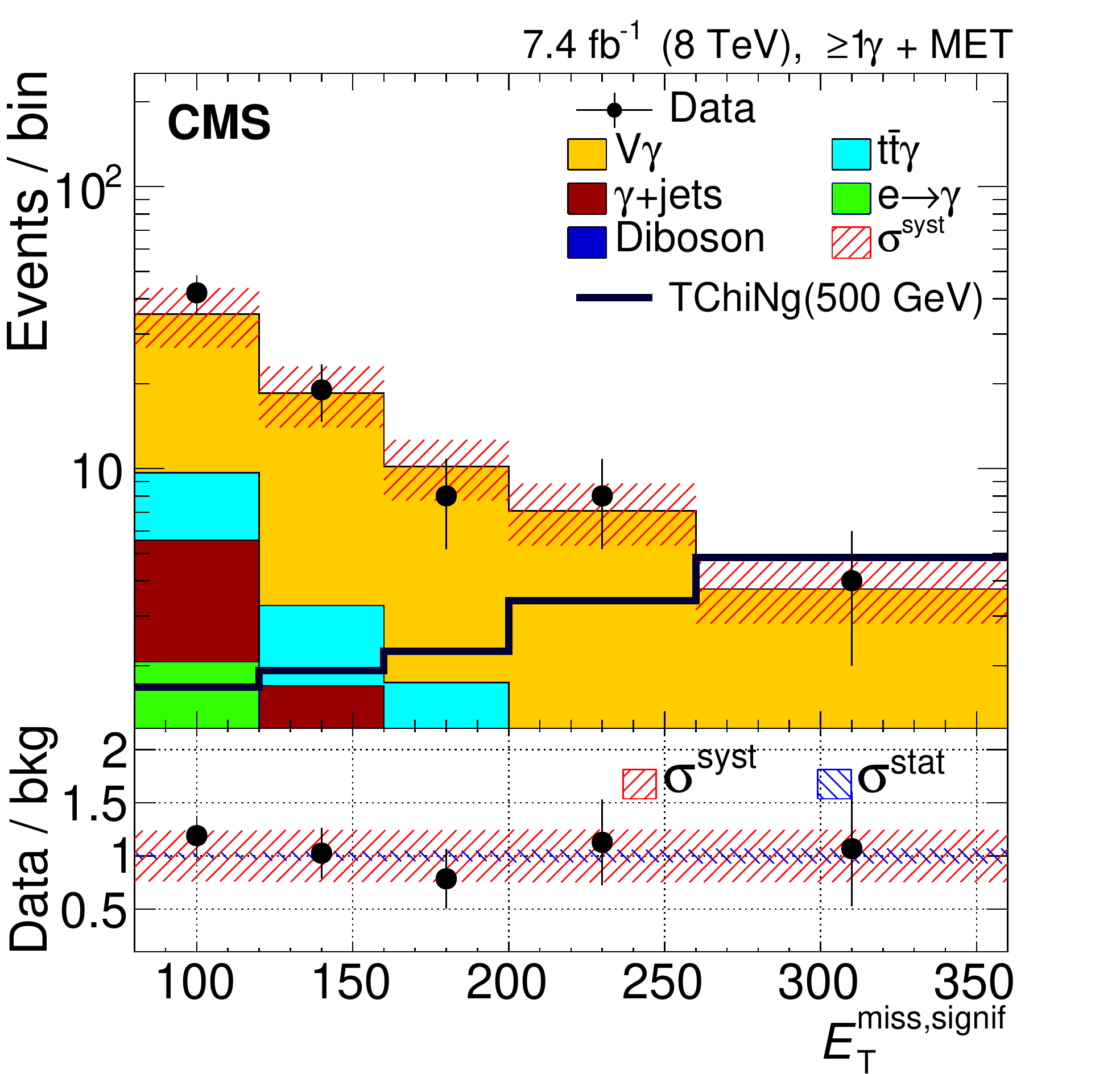}
\includegraphics[width=0.49\textwidth]{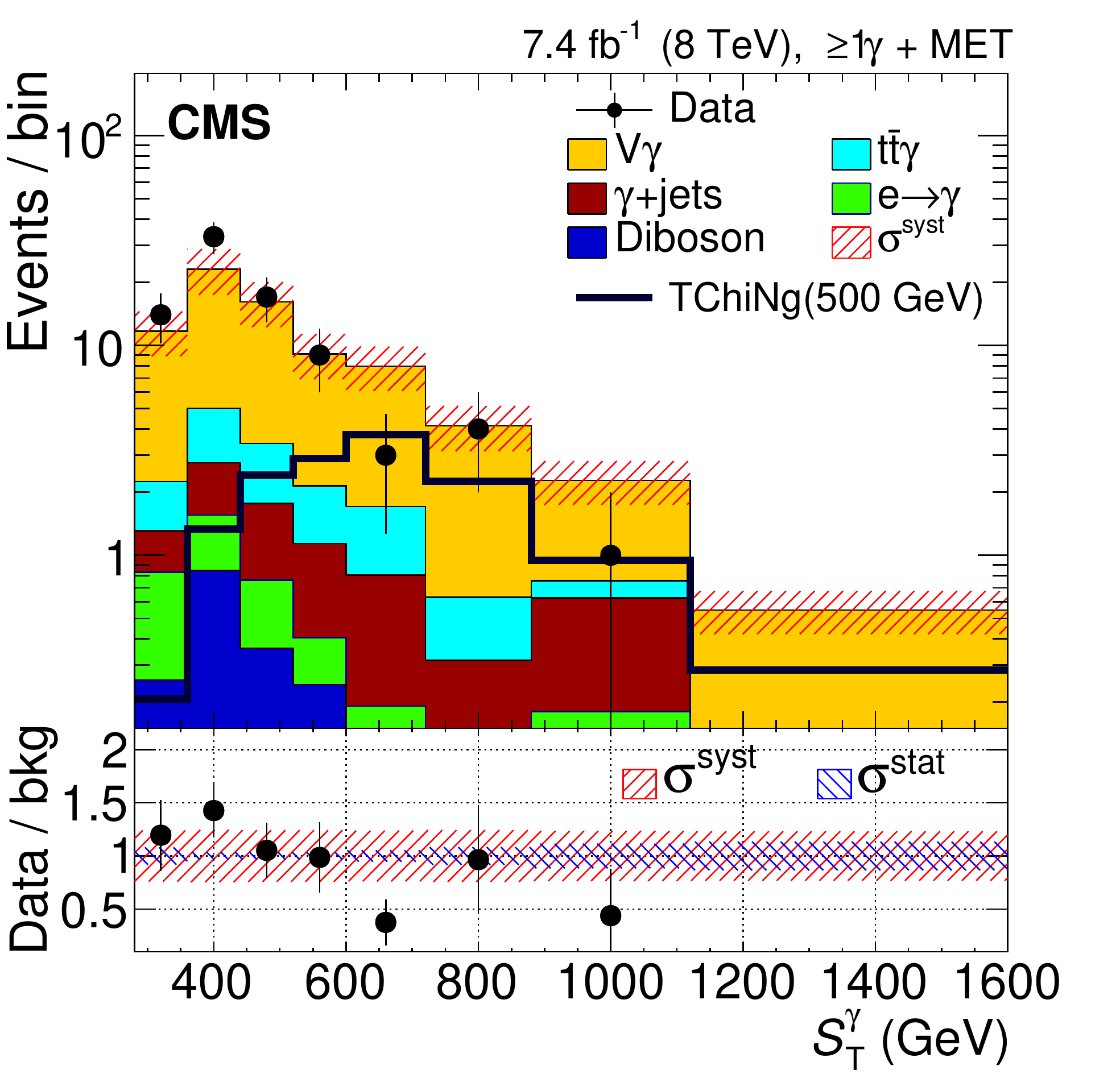}
  \caption{The \metsig (left) and $\JSigma$ (right) variables are shown in the signal selection and used to define four search regions with  \metsig = 200 and $\JSigma$ = 600 GeV partitions. A benchmark TChiNg signal point with an NLSP mass of 500\GeV is shown for comparison.}
 \label{fig:results}
\end{figure*}

As shown in  Fig.~\ref{fig:results}, the observed data are in agreement with the total standard model background expectation within the combined statistical and systematic uncertainties.
Shown are the distributions of \metsig (Fig. 2, left) and \JSigma (Fig. 2, right) used to define the four search regions described in Section 4. The results are summarized in Table~\ref{tab:event_yields}. No sign of new physics is observed.

\begin{table*}[htb]
\centering
\newcommand{\LumiTable}{\ensuremath{7.4\fbinv}\xspace}
\newcommand{\XsecTable}{\ensuremath{\sigma = 6.9\unit{fb}}\xspace}
\newcommand{\ObsLimitTable}{\ensuremath{\sigma_{\mbox{\footnotesize obs}} = 4.9\unit{fb}}\xspace}
\newcommand{\ExpLimitTable}{\ensuremath{\sigma_{\mbox{\footnotesize exp}} = 6.3\unit{fb}}\xspace}
\newcommand{\ObsLimitAsymTable}{\ensuremath{\sigma_{\mbox{\footnotesize obs}} = 5.0\unit{fb}}\xspace}
\newcommand{\ExpLimitAsymTable}{\ensuremath{\sigma_{\mbox{\footnotesize exp}} = 6.4\unit{fb}}\xspace}
\newcommand{\PointTable}{\ensuremath{\text{TChiNg signal point with $M_{\text{wino}}=500\GeV$}}\xspace}
\topcaption{Event yields for data corresponding to \LumiTable and the estimated backgrounds. The signal yields correspond to the benchmark \PointTable shown in Fig.~\ref{fig:results}, also stating the acceptance times efficiency $A\varepsilon$ for each search region. The contribution from QCD multijet background is negligible in all regions.}
\label{tab:event_yields}

\newcommand{\BinLabela}{ \ensuremath{\metsig > 200}, \xspace}
\newcommand{\BinLabelb}{ \ensuremath{\metsig < 200}, \xspace}
\newcommand{\BinLabelc}{ \ensuremath{\metsig < 200}, \xspace}
\newcommand{\BinLabeld}{ \ensuremath{\metsig > 200}, \xspace}
\newcommand{\BinLabelat}{ \ensuremath{\JSigma> 600\GeV} \xspace}
\newcommand{\BinLabelbt}{ \ensuremath{\JSigma> 600\GeV} \xspace}
\newcommand{\BinLabelct}{ \ensuremath{\JSigma< 600\GeV} \xspace}
\newcommand{\BinLabeldt}{ \ensuremath{\JSigma< 600\GeV} \xspace}

\begin{tabular}{rcccc}
\hline
\multirow{2}{*}{Selection}                        &       \BinLabela &       \BinLabelb &        \BinLabelc &       \BinLabeld\\
               &       \BinLabelat &       \BinLabelbt &        \BinLabelct &       \BinLabeldt\\	\hline
$\PV\gamma$               & $\phantom{0}4.7\pm1.2$  & $7.0\pm1.8$ & $42.3\pm10.4$ & $5.0\pm1.3$\\
$\gamma$+jets            & $\phantom{0}0.1\pm0.1$  & $1.3\pm0.3$ & $\phantom{0}3.4\pm0.7\phantom{0}$   & $0.0\pm0.1$\\
$\ttbar\gamma$ & $\phantom{0}0.3\pm0.1$  & $1.1\pm0.3$ & $\phantom{0}5.5\pm1.5\phantom{0}$   & $0.4\pm0.1$\\
Diboson                 & $\phantom{0}0.1\pm0.1$  & $0.2\pm0.1$ & $\phantom{0}1.5\pm0.8\phantom{0}$   & $0.2\pm0.1$\\
$\Pe\to\gamma$    & $\phantom{0}0.1\pm0.1$  & $0.1\pm0.1$ & $\phantom{0}1.6\pm0.2\phantom{0}$   & $0.2\pm0.1$\\
\hline
Background              & $\phantom{0}5.3\pm1.2$  & $9.7\pm1.8$ & $54.3\pm10.6$ & $5.8\pm1.3$\\
Data                    & 4                & 4               & 65                & 8              \\ \hline
Signal                  & $\phantom{0}6.2\pm0.2$  & $2.1\pm0.1$ & $\phantom{0}4.6\pm0.1\phantom{0}$   & $3.3\pm0.1$\\
$A\varepsilon$ [\%] & $12.2\pm0.3$ & $4.2\pm0.1$ & $\phantom{0}9.0\pm0.2\phantom{0}$   & $6.5\pm0.2$\\
\hline
\end{tabular}

\end{table*}

Cross section limits are calculated combining the results of all four search regions defined in the $\JSigma$-$\metsig$ plane at the 95\% CL, using the modified frequentist CL$_{s}$ criterion~\cite{Junk:1999kv,Read,LHCCLs} with a test statistic corresponding to a profile likelihood ratio of the background-only and signal-plus-background hypotheses. Asymptotic formulae~\cite{asymp} are used in the calculation.

\begin{figure*}[tbp]
\includegraphics[width=0.55\textwidth]{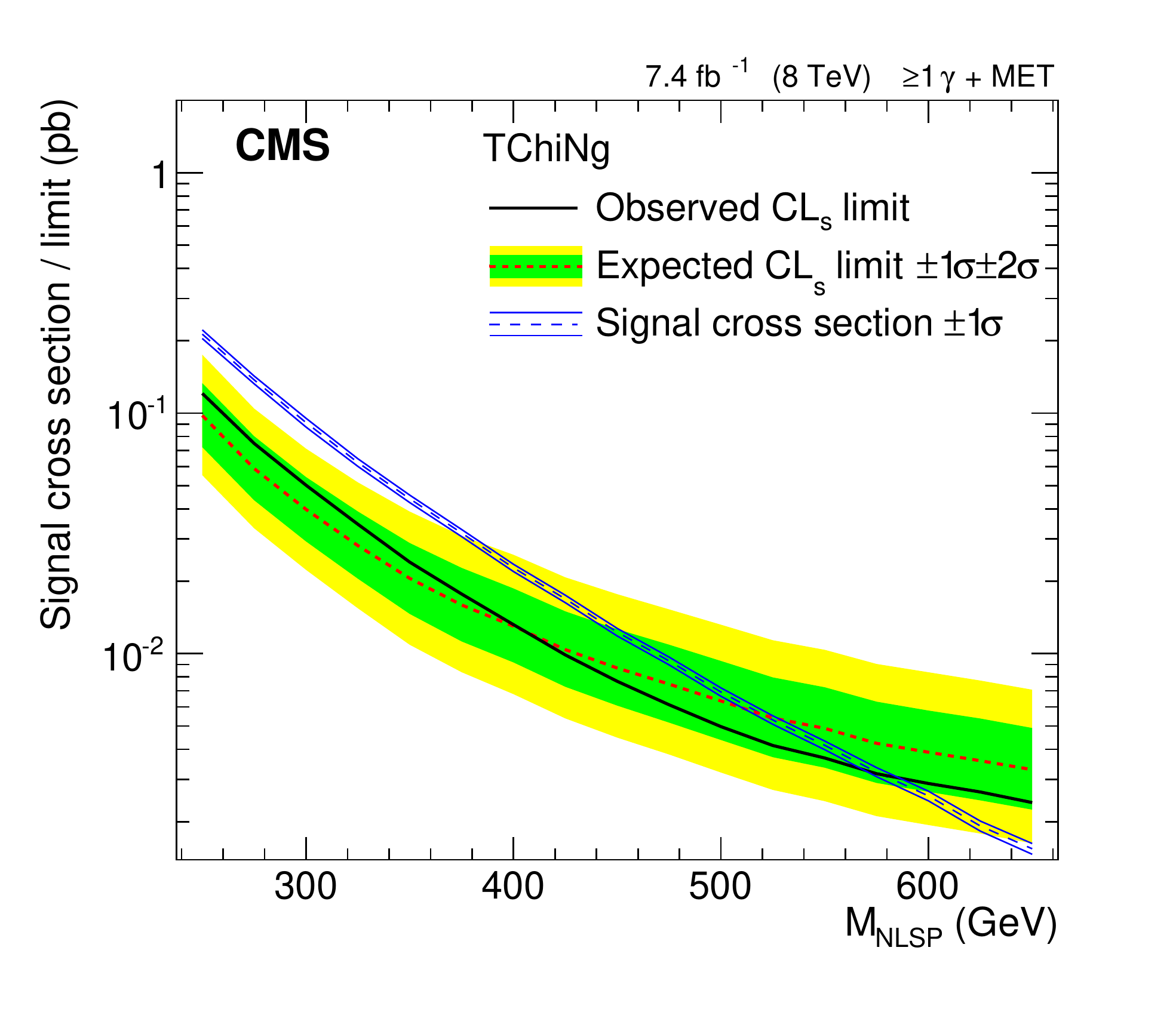}
\includegraphics[width=0.55\textwidth]{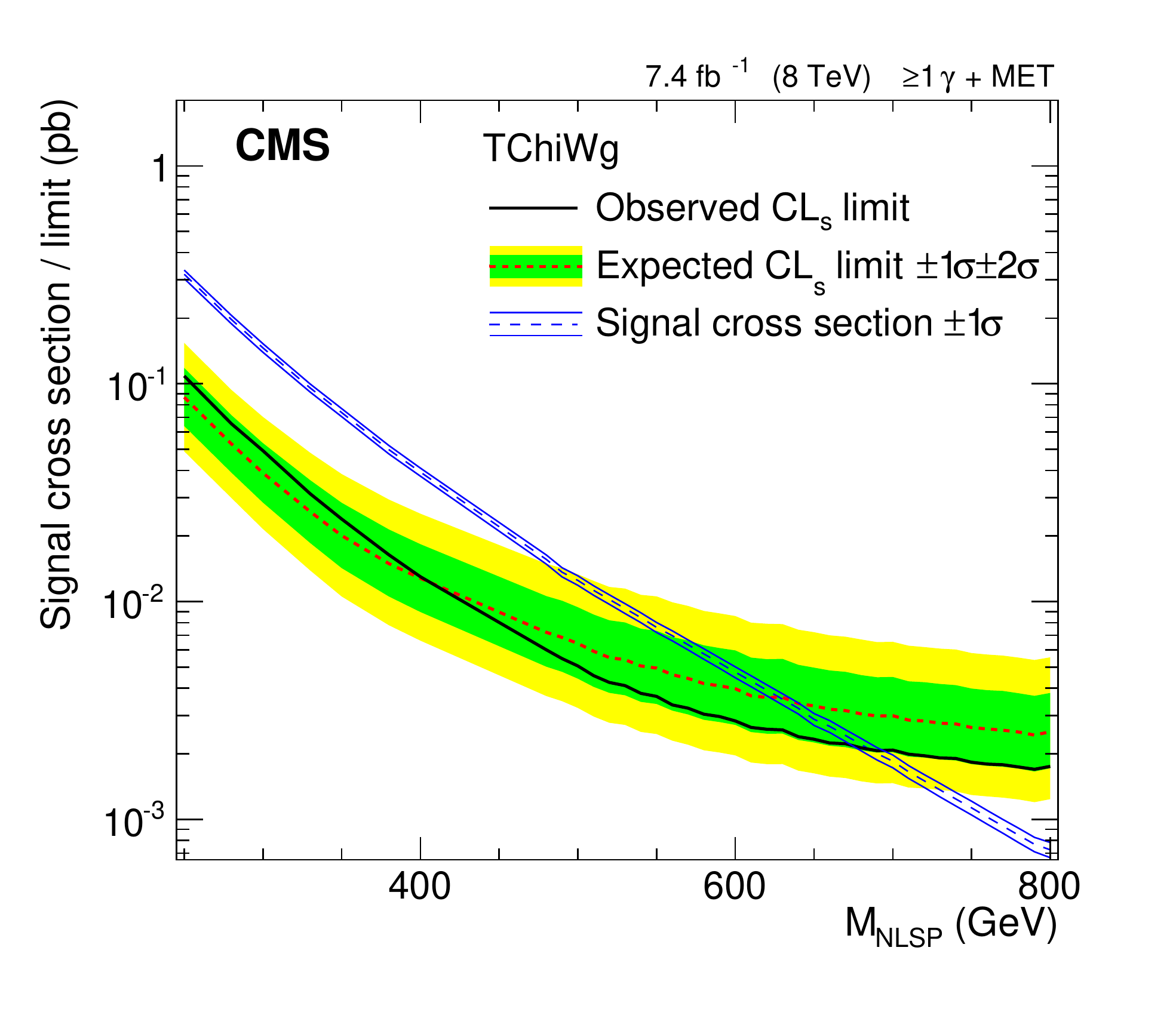}
  \caption{Exclusion limits at 95\%~CL for the TChiNg (left) and TChiWg (right) scenario. In the TChiNg scenario NLSP masses below \TchiNglimit are excluded, in the TChiWg scenario NLSP masses below
	\TchiWglimit are excluded.}
 \label{fig:TChi_exclusion}	
\end{figure*}

The interpretation of the TChiNg and TChiWg scenarios is shown in Fig.~\ref{fig:TChi_exclusion}. The analysis excludes NLSP masses below 570 (680)\GeV at the 95\% CL in the TChiNg (TChiWg) scenario.

\begin{figure*}[tbp]
\includegraphics[width=0.55\textwidth]{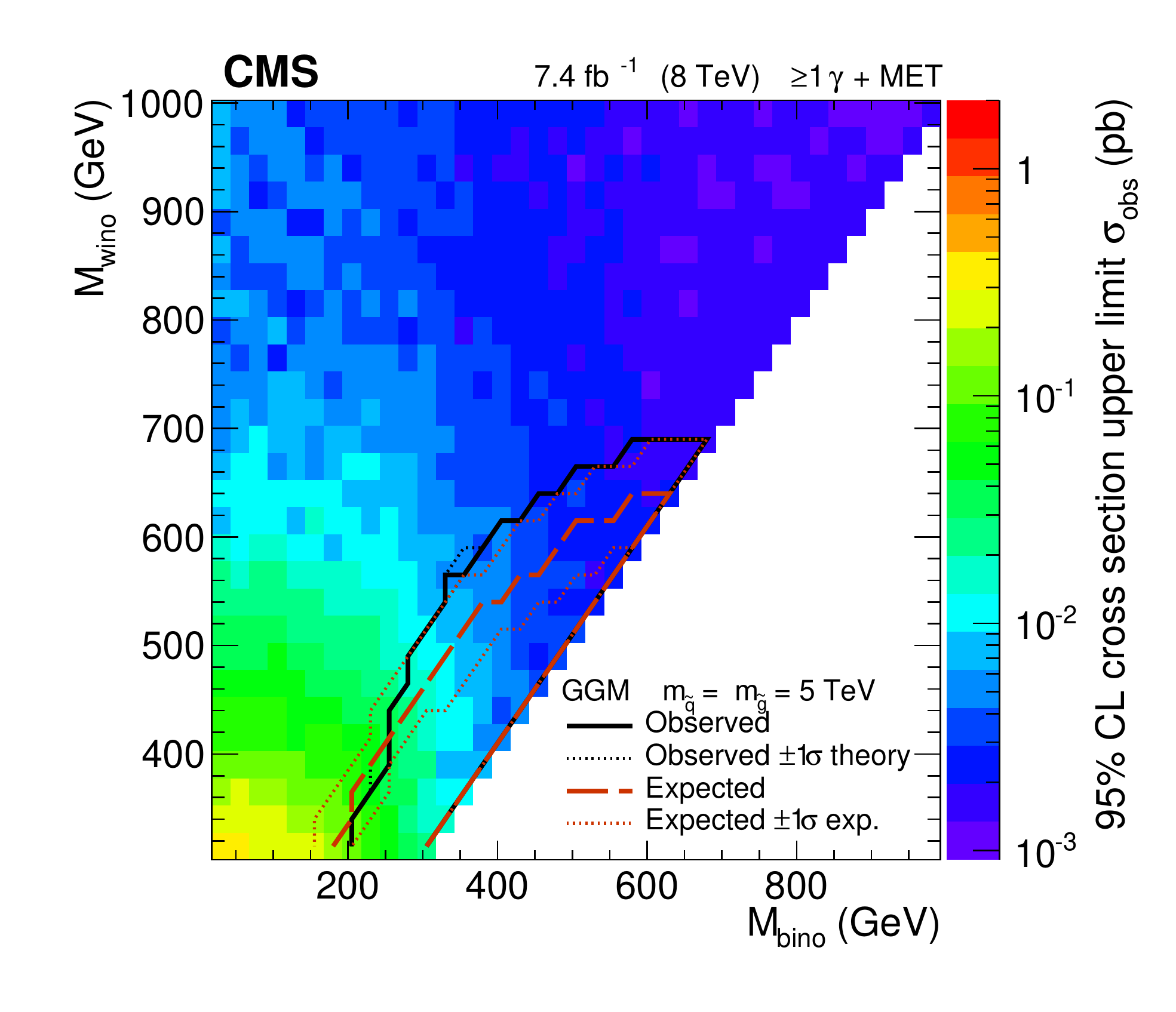}
\includegraphics[width=0.55\textwidth]{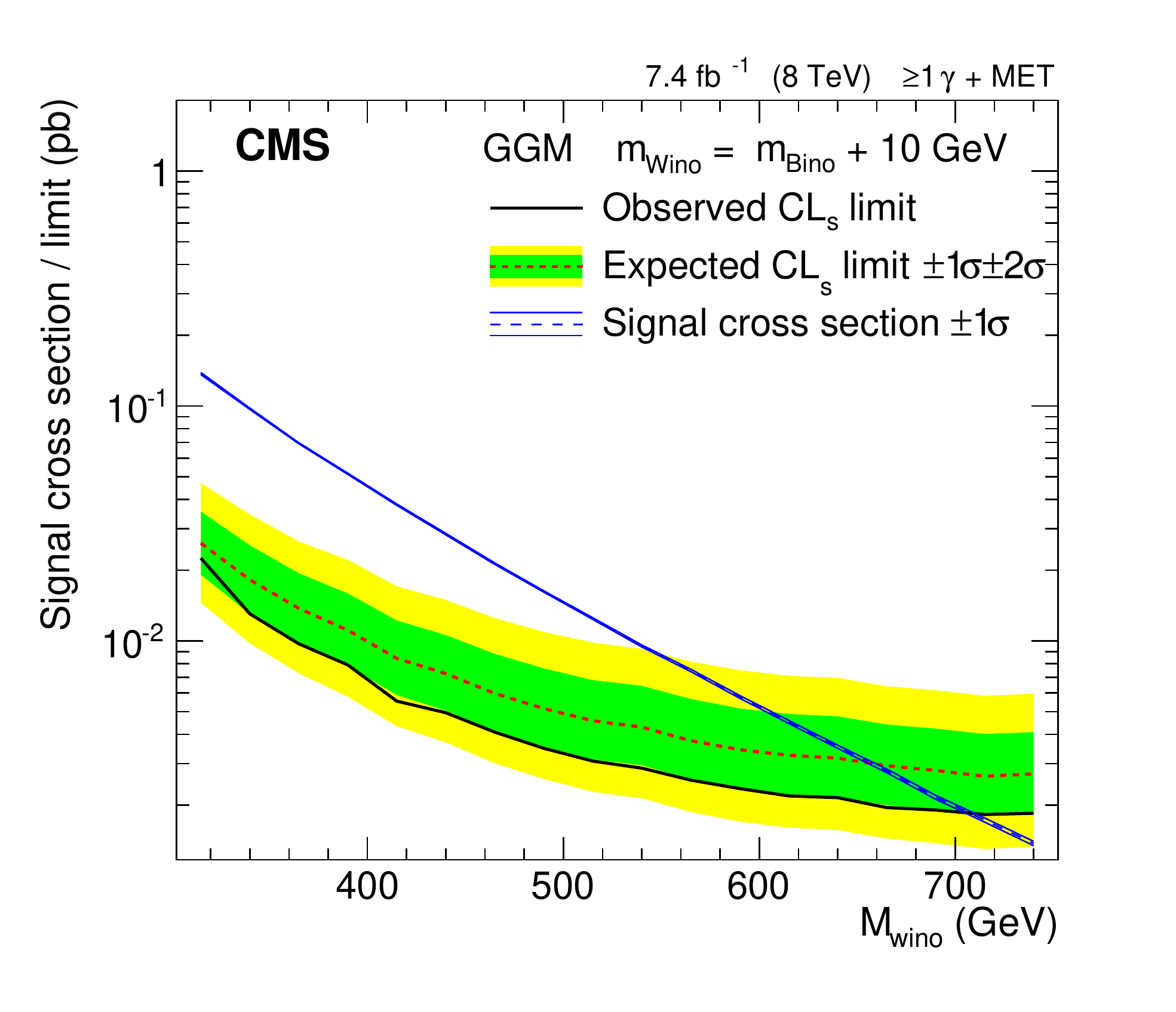}
  \caption{Observed upper cross section CL$_\mathrm{s}$ limit at the 95\%~CL for the GGM signal points in the $M_{\text{wino}}$-$M_{\text{bino}}$ plane (left). Also shown are the 95\%~CL expected and observed exclusion contours. The GGM signal points near the diagonal, e.g. for $M_{\text{wino}} = M_{\text{bino}}10\GeV$ up to a wino mass of $M_{\text{wino}} = \WinoBinoLimit$ are excluded (right).}
 \label{fig:GGM_limits}
\end{figure*}

The 95\%~CL observed upper cross section limit, as well as the observed and expected exclusion contours, for the GGM signal scan in the $M_{\text{wino}}$-$M_{\text{bino}}$ plane are shown in Fig.~\ref{fig:GGM_limits}.
For nearly mass-degenerate gauginos, \ie for $M_{\text{wino}} = M_{\text{bino}}+10\GeV$, wino masses up to approximately $M_{\text{wino}}= \WinoBinoLimit$ are excluded.

\section{Conclusion}
\label{sec:conclusion}

We have searched for electroweak production of gauginos in the framework of gauge mediated supersymmetry breaking in final states with photons and \MET. A dataset, corresponding to an integrated luminosity of \lumiPhoton, recorded with a special trigger with low thresholds is used. The data are found to agree with the SM expectation.
The analysis is sensitive to electroweak production and compressed mass spectra which are characterized by minimal hadronic activity in the final state, complementing previously published searches.
Limits in the TChiNg scenario are set for the first time, excluding NLSP masses below $\TchiNglimit$ at 95\% CL.
In the TChiWg scenario, NLSP masses below $\TchiWglimit$ are excluded at 95\% CL, increasing the previous mass limit in this scenario~\cite{CMS:2015loa} by 140\GeV.
In the general gauge mediation model for compressed mass spectrum scenarios with e.g. $M_{\text{wino}}-M_{\text{bino}} = 10\GeV$, wino masses below  $\WinoBinoLimit$ can be excluded, increasing the previous CMS limit~\cite{Chatrchyan:2012bba} by about 220\GeV.

\begin{acknowledgments}
We congratulate our colleagues in the CERN accelerator departments for the excellent performance of the LHC and thank the technical and administrative staffs at CERN and at other CMS institutes for their contributions to the success of the CMS effort. In addition, we gratefully acknowledge the computing centres and personnel of the Worldwide LHC Computing Grid for delivering so effectively the computing infrastructure essential to our analyses. Finally, we acknowledge the enduring support for the construction and operation of the LHC and the CMS detector provided by the following funding agencies: BMWFW and FWF (Austria); FNRS and FWO (Belgium); CNPq, CAPES, FAPERJ, and FAPESP (Brazil); MES (Bulgaria); CERN; CAS, MoST, and NSFC (China); COLCIENCIAS (Colombia); MSES and CSF (Croatia); RPF (Cyprus); MoER, ERC IUT and ERDF (Estonia); Academy of Finland, MEC, and HIP (Finland); CEA and CNRS/IN2P3 (France); BMBF, DFG, and HGF (Germany); GSRT (Greece); OTKA and NIH (Hungary); DAE and DST (India); IPM (Iran); SFI (Ireland); INFN (Italy); MSIP and NRF (Republic of Korea); LAS (Lithuania); MOE and UM (Malaysia); CINVESTAV, CONACYT, SEP, and UASLP-FAI (Mexico); MBIE (New Zealand); PAEC (Pakistan); MSHE and NSC (Poland); FCT (Portugal); JINR (Dubna); MON, RosAtom, RAS and RFBR (Russia); MESTD (Serbia); SEIDI and CPAN (Spain); Swiss Funding Agencies (Switzerland); MST (Taipei); ThEPCenter, IPST, STAR and NSTDA (Thailand); TUBITAK and TAEK (Turkey); NASU and SFFR (Ukraine); STFC (United Kingdom); DOE and NSF (USA).
\end{acknowledgments}
\ifthenelse{\boolean{cms@external}}{\vspace*{2em}}{} 

\bibliography{auto_generated}

\cleardoublepage \appendix\section{The CMS Collaboration \label{app:collab}}\begin{sloppypar}\hyphenpenalty=5000\widowpenalty=500\clubpenalty=5000\textbf{Yerevan Physics Institute,  Yerevan,  Armenia}\\*[0pt]
V.~Khachatryan, A.M.~Sirunyan, A.~Tumasyan
\vskip\cmsinstskip
\textbf{Institut f\"{u}r Hochenergiephysik der OeAW,  Wien,  Austria}\\*[0pt]
W.~Adam, E.~Asilar, T.~Bergauer, J.~Brandstetter, E.~Brondolin, M.~Dragicevic, J.~Er\"{o}, M.~Flechl, M.~Friedl, R.~Fr\"{u}hwirth\cmsAuthorMark{1}, V.M.~Ghete, C.~Hartl, N.~H\"{o}rmann, J.~Hrubec, M.~Jeitler\cmsAuthorMark{1}, V.~Kn\"{u}nz, A.~K\"{o}nig, M.~Krammer\cmsAuthorMark{1}, I.~Kr\"{a}tschmer, D.~Liko, T.~Matsushita, I.~Mikulec, D.~Rabady\cmsAuthorMark{2}, N.~Rad, B.~Rahbaran, H.~Rohringer, J.~Schieck\cmsAuthorMark{1}, R.~Sch\"{o}fbeck, J.~Strauss, W.~Treberer-Treberspurg, W.~Waltenberger, C.-E.~Wulz\cmsAuthorMark{1}
% ...
\vskip\cmsinstskip
% ...
1:~~Also at Vienna University of Technology, Vienna, Austria\\
2:~~Also at CERN, European Organization for Nuclear Research, Geneva, Switzerland\\
% ...

\end{sloppypar}
\end{document}